# Manufacturing and testing a thin glass mirror shell with piezoelectric active control

D. Spiga[1][§], M. Barbera[2], A. Collura[3], S. Basso[1],
R. Candia[3], M. Civitani[1], M. Di Bella[3], G. Di Cicca[3], U. Lo Cicero[3],
G. Lullo[2], C. Pelliciari[1], M. Riva[1], B. Salmaso[1], L. Sciortino[2], S. Varisco[3]

[1]INAF – Osservatorio Astronomico di Brera, Via Bianchi 46, 23807 Merate (Italy)
[2]Università degli Studi di Palermo, Via Archirafi 36, 90123 Palermo (Italy)
[3]INAF - Osservatorio Astronomico di Palermo, Piazza del Parlamento 1, 90134 Palermo (Italy)

## ABSTRACT

Optics for future X-ray telescopes will be characterized by very large aperture and focal length, and will be made of lightweight materials like glass or silicon in order to keep the total mass within acceptable limits. Optical modules based on thin slumped glass foils are being developed at various institutes, aiming at improving the angular resolution to a few arcsec HEW. Thin mirrors are prone to deform, so they require a careful integration to avoid deformations and even correct forming errors. On the other hand, this offers the opportunity to actively correct the residual deformation: a viable possibility to improve the mirror figure is the application of piezoelectric actuators onto the non-optical side of the mirrors, and several groups are already at work on this approach. The concept we are developing consists of actively integrating thin glass foils with piezoelectric patches, fed by voltages driven by the feedback provided by X-rays. The actuators are commercial components, while the tension signals are carried by a printed circuit obtained by photolithography, and the driving electronic is a multi-channel low power consumption voltage supply developed in-house. Finally, the shape detection and the consequent voltage signal to be provided to the piezoelectric array are determined in X-rays, in intra-focal setup at the XACT facility at INAF/OAPA. In this work, we describe the manufacturing steps to obtain a first active mirror prototype and the very first test performed in X-rays.

## 1. INTRODUCTION

The angular resolution of X-ray telescopes is an essential figure of merit for describing the capabilities for detecting X-ray sources in the distant Universe. However, also large mirror apertures are required to reach large effective areas and the high sensitivity needed to resolve faint X-ray astronomical sources. Manufacturing large, lightweight, and high angular resolution optical modules will represent the challenge of the next years for astronomical X-ray optics: the current requirements for e.g., the ATHENA telescope (currently selected by ESA for launch in 2028) are a 2 m$^2$ effective area at 1 keV with an angular resolution of 5 arcsec HEW (Half Energy Width), obtained filling a 3 m diameter module with high-resolution modular elements (XOU, X-ray Optical Units) having a common focus. The large dimensions of the optics and the dense mirror nesting needed to meet the effective area requirements entail the adoption of lightweight materials such as glass or silicon; for ATHENA, the baseline technology adopted is Silicon Pore Optics[1] (SPO), developed by ESA/ESTEC and Cosine since 2004, and still ongoing.

However, Slumped Glass Optics (SGO) represents a viable alternative to manufacture high-resolution, lightweight and densely nested optical modules[2]. In this approach, already experienced for the NuSTAR telescope[3], thin foils of glass are formed at high temperature onto a precisely figured mould and, kept at the correct distance by ribs, stacked into an XOU. The ribs also provide the mechanical stiffness to the XOU needed to maintain its shape during the operation. For the ATHENA (formerly IXO) X-ray telescope, the SGO technology has been developed at INAF/OAB under ESA/ESTEC contract in 2009-2013, and in parallel at MPE (Garching, Germany), following two different methodologies (*direct* slumping at OAB[4] and *indirect* slumping at MPE[5]). In both cases, as well as for SPOs, it was proven that an optical design based on SGO was able to provide the requested effective area/mass ratio.

---

[§] contact author: daniele.spiga@brera.inaf.it, phone +39-02-72320427

Currently, efforts are concentrated on improving the angular resolution of optical modules. It is well known that the degradation of the imaging properties of X-ray mirrors may result from two factors: one is the surface microroughness that triggers a reduction of a reflectivity in the focus, together with an increase of the X-ray scattering (XRS) in the nearby directions and consequent damaging of the angular resolution. This effect exhibits a marked increase for increasing energy of the X-rays, and has to be kept to tolerable level by fine polishing of the surface. Glass foils with the required[6] thermal properties, thickness (< 0.4 mm), and surface smoothness (rms < 4 Å over in the 2 mm – 20 µm range of lateral spatial wavelengths) are commercially available, such as D263T and AF32 by Schott, or EAGLE-XG by Corning. In the direct slumping approach, the optical surface comes directly to contact with the forming mould; hence, it is completely unaffected by non-uniformities in the glass foil thickness. However, the surface roughness of the glass foils can be damaged by the shear on the forming mould, as a consequence of the different Coefficient of Thermal Expansion (CTE) of the two materials. By an appropriate choice of the mould material (Zerodur K20) and a process optimization via a dedicated study carried out at INAF/OAB, this difference has been reduced enough to preserve the surface smoothness to acceptable levels (less than 1 arcsec of HEW degradation caused by XRS at 1 keV).

The other source of imaging degradation is represented by deformations of the optical surface. In the direct slumping methodology, the longitudinal profile of the X-ray mirror is subject to non-negligible errors that have been continuously reduced[6] by a proper optimization of the slumping setup (mould height, intensity and duration of the pressure exerted, surface cleanliness, settings of the slumping cycle...), and partly corrected by the development of an integration process[4], which enforces the slumped foils with the Wolter-I profile of accurately figured integration moulds. Fixing the stack on the glass via stiffening glass ribs, the nominal shape is imparted to the optical surface at the rib locations. Because of the glass springback, however, the correction is only partial especially for defects ranging on a few centimeters of lateral scale in the longitudinal direction. The residual profile error is also a function of the distance from the nearest rib: more exactly, it is maximum mid-way between adjacent ribs, while only at the rib location the profile copies almost exactly the nominal shape imparted by the integration mould. As a result, the expected angular resolutions of slumped glass mirrors are approaching the target of 5 arcsec HEW, but are still far from the sub-arcsec resolutions envisaged for some mission concepts[7],[8]. Even if further process improvements are surely possible, several groups worldwide are at work on the possibility to actively correct the mirror profile after forming. However, the dense nesting, typical of X-ray optics, does not permit using normal actuators, as commonly done in optical astronomy. In contrast, using piezoelectric elements acting *tangentially*, i.e., exerting a strain in a direction parallel to the surface on the rear of the glass foil, the local curvature of the mirror can be changed and profile errors corrected.

In this paper we describe last year's activities carried out in the AXYOM (*Adjustable X-raY optics for astrOnoMy*) project involving INAF/OAB, Università di Palermo, and INAF/OAPA, to study the possibility to actively correct the shape of thin glass X-ray mirrors[9]. While other groups[8] have selected the approach of depositing patches of piezoelectric thin films on slumped glassed, we have adopted piezoelectric elements available commercially, to be fed by a system of metallic electrodes deposited by photolithography on the non-optical side of the foil. In particular, piezoceramic patches were selected because they are characterized by a sufficient strength to bend the glass foil even with moderate electrical voltages[10]. A dedicated electronic board supplies the required voltages to the piezoelectric elements. Finally, the feedback to the piezoelectric array should be provided by reconstructing the mirror shape under X-ray illumination in intra-focal setup[11] at the XACT facility[12],[13] at INAF/OAPA, following an algorithm initially developed to solve beam-shaping problems[14]. The AXYOM project naturally inscribes in the experience gained at the participating institutes in hot slumping, integration of thin glass foils, thin film deposition, photolithography, electronics, and X-ray tests. Moreover, facilities are already available to achieve each of these tasks and produce active mirrors at a very low cost. We succinctly describe in this paper the steps taken to produce a first active mirror prototype with two small piezoelectric patches, solely aimed at testing the manufacturing procedures and the experimental X-ray setup. A more detailed description of the mirror manufacturing process is reported in another paper[15]. The realization of a more advanced prototype is foreseen for late November 2015.

## 2. THE MIRROR FABRICATION

The first step in the production consists of preparing a formed glass with the electrical contacts that have to bring the voltages to the piezoelectric elements: the electrode design is obviously a function of the size and the location of the piezo actuators. The model that has been selected is a flexible piezoceramic transducer sold by *Physik Instrumente,* mod. P-876.SP1, with a 16 mm × 16 mm size, and a thickness of only 200 µm to avoid X-ray obstruction in a mirror stack. The piezo transducers are provided enclosed in Kapton, with two soldering pads on the same side, can be freely used for high-vacuum applications, and operate in a voltage range from -100 V to +400 V, much larger than the values expected for a correction of mirror defects up to a maximum amplitude of a few microns. In this preliminary design, the piezoelectric elements are located where the forming error amplitude is expectedly larger, i.e., mid-way between ribs (see Sect. 1). The electrodes have also to be very thin to avoid not only obstruction of the X-ray flux, but also to avoid a mirror deformation by their weight. To this end, they are printed by photolithography on the backside of the mirror: they are structured as thin metallic films (15 nm of titanium and 80 nm of gold) and, since they have to supply static tensions to the actuators (that electrically behave as capacitors), their conductivity does not need to be very high. The thin printed circuits (Fig. 1) brings voltages from a matrix of 5 × 5 bipolar slots to one lateral side of the glass, where the printed tracks can be interfaced to the feeding electronics. The weight of the contact wires clearly deforms one of the mirror sides, but, owing to the rib stiffness, the deformation is limited to the region between the wire junctions and the nearest rib. The broader tracks are common contacts and serve as ground electrode.

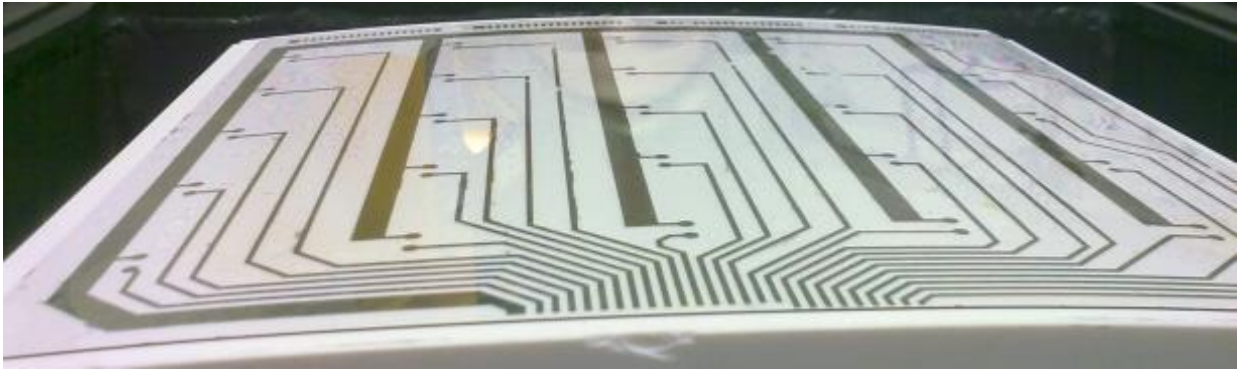

Fig. 1: system of metallic electrodes deposited on the backside of an uncoated slumped glass, laid upon the K20 slumping mould. The azimuthal curvature, previously imparted by the hot slumping process, can be noticed.

The electrodes were deposited onto a 0.4 mm thick EAGLE-XG glass foil previously slumped at INAF/OAB to cylindrical shape with a 200 mm × 200 mm size and 1 m radius of curvature. The photolithographic process has been studied at INAF/OAPA, and – among different possibilities – the *chemical attack* method was selected, consisting of the following steps:

1) the titanium and gold layers are evaporated onto the non-optical side of the slumped glass;
2) a photoresist layer is sprayed on the metallic film;
3) the mask is printed in positive and laid onto the photoresist;
4) the mask is exposed to UV rays, shading the photoresist where the metal has to be left;
5) the photoresist is developed with NaOH, leaving the metal exposed where the mask was not present;
6) the gold and titanium layers outside the printed tracks are removed in baths of aqua regia and a $H_2O_2$ solution;
7) the residuals of photoresist are washed out by a NaOH solution.

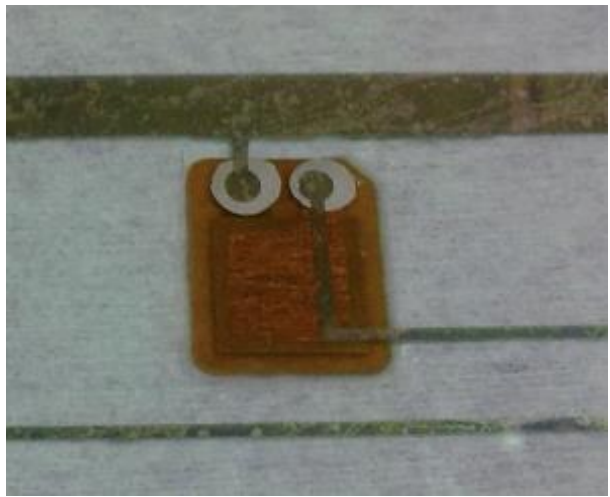

Fig. 2: (left) a detail of the glued piezoelectric patch seen through an uncoated glass foil. The electrical contacts are ensured via two droplets of conductive epoxy.

This methodology has shown excellent results in terms of contact integrity. The tracks are conductive and exhibit only moderate deformation (0.2 µm of P-V over lateral scales of 4 cm) at the grounded contacts; to avoid this problem, the next prototypes will be designed with the same width of all the tracks. The glass was coated with an 80 nm titanium layer to suppress the reflection of the backside of the glass in visible light. This is necessary to enable metrological characterizations with optical sensors, such as the Long Trace Profilometer (LTP) used to preliminarily characterize the influence functions (Fig. 3, right).

For this test, aimed at investigating the criticalities of the mirror production and of the experimental setup at XACT, only two piezoceramic elements were used. They were glued at the contact terminals in the central sector using a thin layer (75 µm) of low-shrinkage epoxy glue (Masterbond EP30-2), the same routinely used for the integration of slumped glass optics at OAB. To ensure the electrical contacts, two droplets of conductive glue were used to connect the soldering pads (Fig. 2), paying attention to avoid short-circuits. As this was the first test, the piezoelectric patches were *not* integrated forcing the glass in contact with the mould by vacuum suction (as the standard integration procedure would have required). Therefore, we may expect a local deformation induced by the glue much larger than achievable with the next tests. The electrodes connected to the piezos were fixed to electrical wires using the same conductive glue. The wires are used to connect a power supply or the electronic board (Sect. 3).

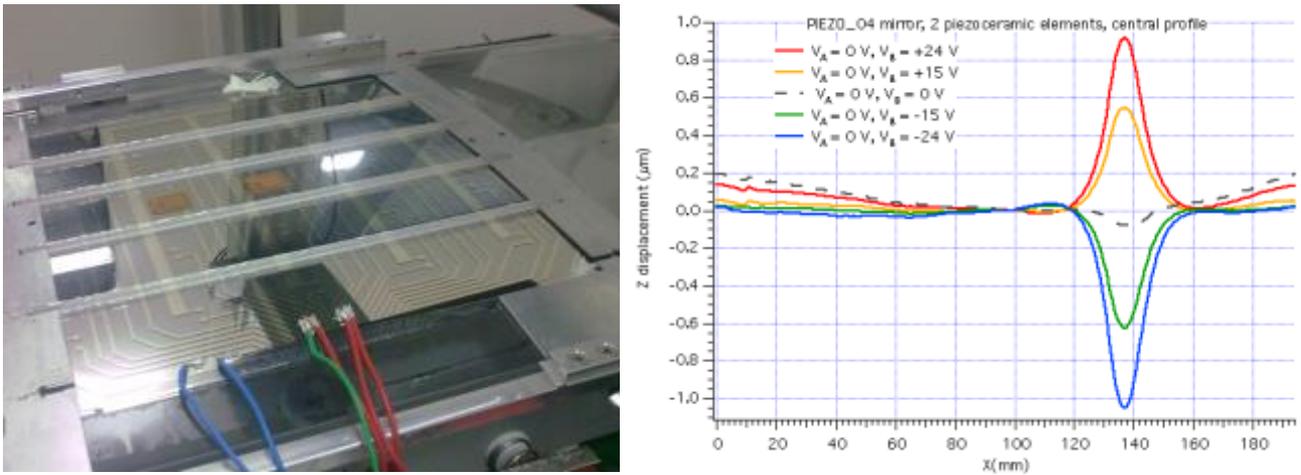

Fig. 3: (left) the active mirror being integrated: two piezoelectric patches are glued on the electrical contacts. The corresponding contacts are connected to wires in order to supply the necessary voltages. The blue cables are connected to ground. Also 4 ribs are being glued, aligned by precision masks. (right) influence functions of the integrated foil, measured at the Long Trace Profilometer at INAF/OAB, varying the voltage provided to the piezo B.

After the complete glue polymerization, the actuated glass was finally integrated onto an aluminum backplane at OAB using the IMA (Integration Machine[4]) developed for SGO stacking (Fig. 3, left). Four ribs in BK7 glass were used to support the glass onto the backplane. Because the glass and the aluminum have different CTEs, and the integration is performed at the temperature of $(20 \pm 0.2)$°C, it is very important the temperature of the testing chamber at XACT (Sect. 4) to remain as close as possible to this value to avoid deformation or even breakage of the glass. The longitudinal profile of the integrated module has been measured at the LTP to check the operation of the piezoelectric elements when different voltages are applied in the range -25 V to +25 V. The net displacement in the *z* direction when the piezo B is fed is shown in Fig. 3, right, proving the capability to perform the correction of profile errors in a vertical range of a few microns.

## 3. THE MULTICHANNEL PIEZO DRIVER

A dedicated electronic circuit (Fig. 4) has been fabricated at UNIPA for driving the piezo actuators applied to the active mirror. At this stage of the project, attention has been focused more on circuit flexibility rather than ultimate optimization. As a consequence, circuit miniaturization and ultra low power consumption have been traded off versus other parameters, such as scalability of the number of channels, circuit simplicity and stability, maximum output voltage and power supply efficiency. After a number of theoretical considerations, circuit simulations and tests, the prototypal system has been developed as follows.

The signals controlling the multichannel piezo driver are generated by a commercial 16 channels digital to analog converter, model USB-3105, manufactured by Measurement Computing Corporation. Each channel of the converter can generate a voltage between -10 V and +10 V with a 16-bit resolution. The converter is controlled and powered by a personal computer through an USB connection. The amplitude of each control signal is then increased by a voltage amplifier, to obtain a maximum output voltage span of ±70 V. Two boards have been designed and fabricated, each allocating eight amplifiers, stacked together to reach the number of 16 channels per system.

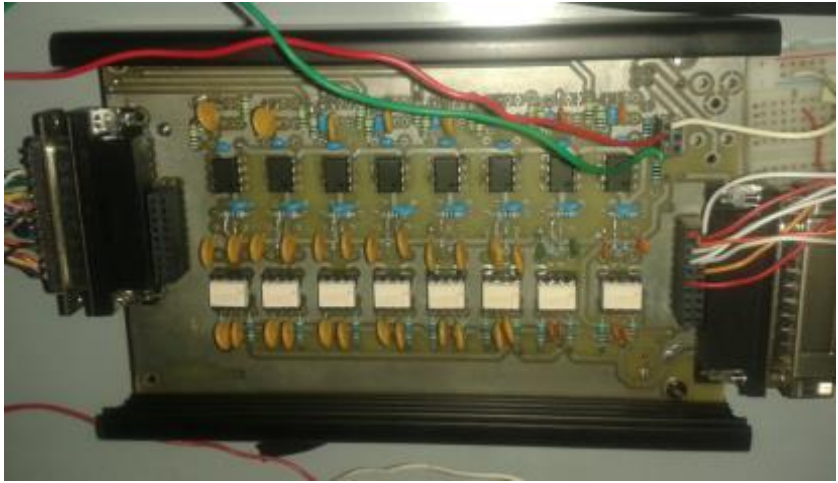

Fig. 4: one of the boards developed to control eight piezoelectric elements. Eight aligned amplifiers are visible.

Finding integrated amplifiers capable of managing the specified output voltages has proved to be quite difficult. A number of high voltage monolithic operational amplifiers (OP AMP) had been considered during the design, but they had to be ruled out due to their high power consumption. The typical quiescent current for such an amplifier is around 2.5 mA per channel, even with no applied input signal. Considering a ±70 V dual supply, it means a power dissipation of 0.35 W per channel.

As a consequence, a multichannel high voltage and low dissipation voltage amplifier has been designed to the purpose. Each channel is composed by two stages. The first stage is a low power OP AMP, powered by a ±12 V dual power supply, which drives the high voltage output stage. In conventional amplifier design, the output stage is typically an emitter voltage follower obtained using a couple of complementary transistors. As a consequence, the driver stage needs to supply the full span of the desired output voltage. In our design, a different approach has been used. The high voltage output stage consists of a pair of identical optocouplers, model TLP627 manufactured by Toshiba, whose output transistors are connected in series between the ±70 V dual supply voltages. It recalls a "totem pole" configuration, where the two optocouplers act as pull-up and pull-down devices. The LED's inside the optocouplers are driven in a push-pull configuration by the low voltage OP AMP. The output stage can thus be considered as a current amplifier, with a minimum current gain of 10 and capable of managing high voltages, driven by the low voltage OP AMP. The voltage feedback, derived from the output voltage, guarantees the desired signal amplification. The amplifier has an overall voltage amplification of 10, with a bandwidth limited to 100 Hz, well compatible with the typical control signals driving the piezo actuators.

There are many advantages in the proposed configuration. The OP AMP driving stage operates at low voltages, thus decreasing the quiescent power consumption. Furthermore, the output stage, based on optocouplers, has very low quiescent power consumption, given by the small bias current (around 0.1 mA) needed to increase the output stage linearity. In our circuit, power consumption is thus reduced to less than 15 mW per channel. Finally, the adopted optocouplers have a maximum operation voltage of 300 V. It means that the output stage is capable of managing voltages between -150 V and +150 V. As in this project we were interested, for driving the piezo actuators, in a voltage span of only ±70 V, the voltages generated by the power supply were limited to this range.

As for the "high voltage" power supply stage, it is based on the LT8300 integrated circuit, a micropower isolated flyback converter produced by Linear Technology. Using a dual output transformer with a 1:5 winding ratio, it is capable of producing a ±70 V dual output voltage with a typical maximum output current of 4 mA, starting from a single 12 V supply. It means that the whole 16 channels piezo driver only needs a ±12 V supply. Thanks to the adopted strategies, the total power consumption for the 16 channels voltage amplifier has been measured to be within 0.5 W.

# 4. FIRST TESTS AT THE XACT FACILITY

The 35 m-long XACT facility at INAF/OAPA is already described in detail elsewhere[12][13] and the analysis of the intra-focus intensity distribution pattern, which should enable us to reconstruct the mirror profile and to find the optimal voltages to perform the correction of the figure errors, is duly exposed in a previous paper[11]. In the present case, however, since the mirror had only two small piezoelectric actuators, to be compared with a total mirror surface of 200 mm × 200 mm, the correction capabilities were extremely limited. The first tests performed at the XACT (Fig. 5, left) in July 2015 were not addressed at an effective correction of the mirror profile, but rather to a test of the equipment and of the experimental setup at the XACT facility. In particular, we could successfully test the mechanical interfaces of the backplane to the alt-azimuth manipulator (Fig. 5, right), the electrical connection of the piezos to the electronic board (of which only two channels were used). We could also experiment a thermo-stabilization system of the vacuum chamber using an external circulation of cooled water, which succeeded to bring the mirror temperature to 21.5 °C, i.e. quite close to the temperature at which the mirror was integrated.

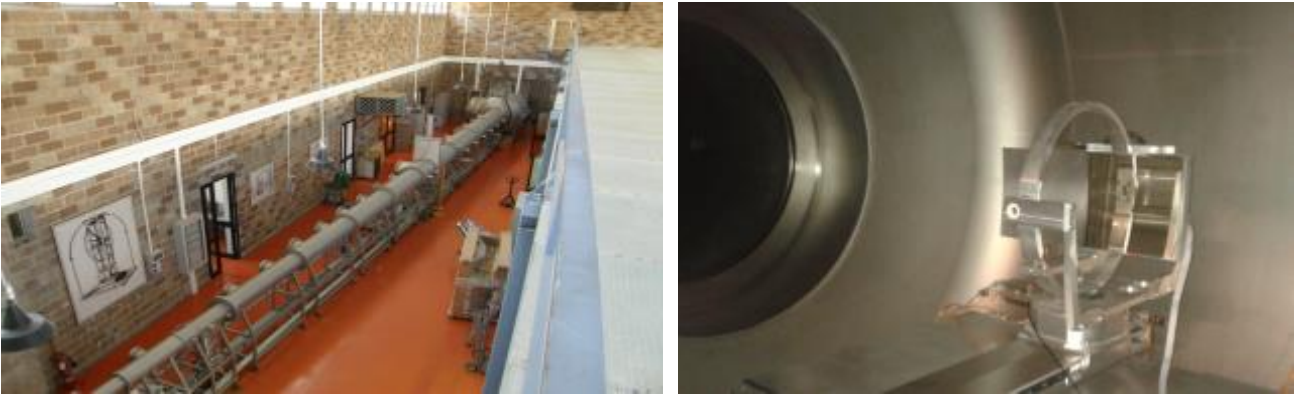

Fig. 5: (left) a view of the XACT X-ray facility at INAF/OAPA. (right) the integrated mirror installed in the manipulation chamber of XACT. Rays impinge from the right side of the image, at an angle of approximately 2.2 deg. The porthole visible on the left side of the picture leads to the microchannel plate detector, at a 4965 mm distance. The module under test is mounted on a translation-rotation manipulator stage, with the optical surface of the mirror at a 210 mm distance from the central axis.

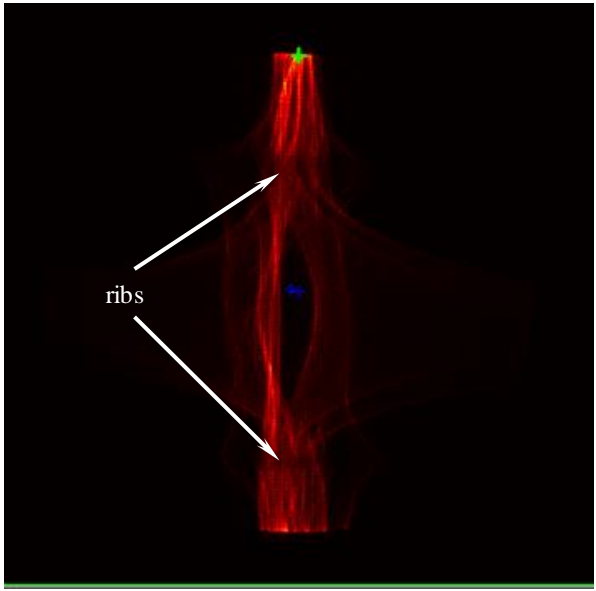
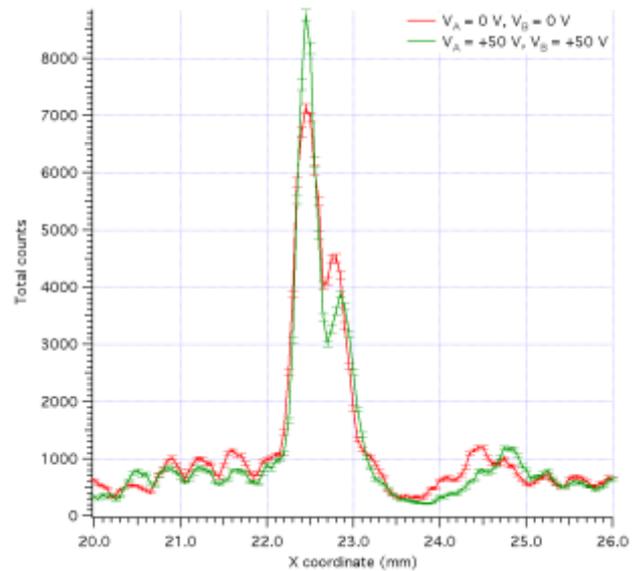

Fig. 6: (left) the central part of the intrafocal trace of the active mirror seen in the 4 cm diam. field of the MCP at XACT at the X-ray energy of 0.3 keV (the C-K$\alpha$ fluorescence line). The location of the two central ribs is clearly seen. The central part is the one where the piezo patches were glued. (right) variation of the intensity distribution for two example voltages provided to the piezo A and B. Also statistical error bars are shown.

After alignment using a laser mounted in the 35 m tube, the chamber was evacuated and the mirror illuminated with X-rays at the energy of 0.3 keV. The value was selected to be as small as possible in order to minimize the XRS caused by surface roughness, which would make the image completely confused, hindering any possibility to reconstruct the mirror profile. The piezoelectric elements survived in vacuum (as it could be checked measuring the capacitances at the contacts outside the chamber) and no degassing problem was reported. The intra-focal trace appeared clearly in the 4 cm diameter microchannel plate (MCP) with a 0.1 mm resolution. No visible scattering effects were observed.

The central part of the intra-focal trace amid ribs is the region where the piezos have been fixed (Fig. 6, left). As anticipated in Sect. 2, there is a signature of deformation, probably caused by gluing done without keeping the glass in contact with the integration mould. This fast and simplified gluing was done only as an initial test. The intensity distribution is quite difficult to understand because the intensity distribution lines cross each other, making the profile reconstruction complicated. Anyway, the effect of the piezo actuation can seen because the reflected trace intensity changes distribution when the piezoceramic elements are fed by a voltage in the range -50 V /50 V (Fig. 6, right). Although this does not allow us performing a shape correction, it demonstrates that the mirror deformation imparted to the optical surface can be changed from outside the chamber acting on the voltages and that we can see the effects directly in X-ray illumination.

## 5. FUTURE ACTIVITIES

The experiment shown in this paper was only a first test - in the context of the AXYOM project - of the manufacturing procedures studied to obtain an active X-ray mirror using commercial piezoelectric components, and of the possibility to detect the shape changes in X-ray illumination, providing in principle the feedback to the voltage controls.

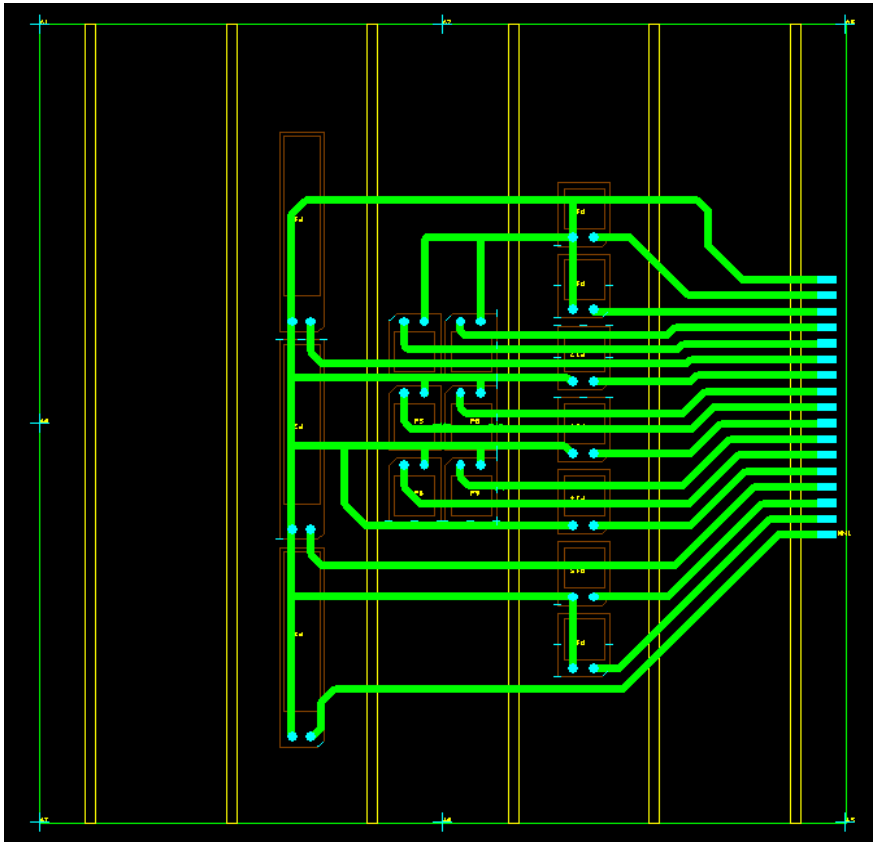

Fig. 7: design of the next electrode pattern to be realized. Different distributions of the piezoelectric elements will be adopted in different sections between ribs. Also longer piezoelectric elements (P-876K015) will be used to actuate a longer surface with a lower number of contacts.

The next test at the XACT facility is foreseen for late November 2015. Taking advantage of the lessons learnt in this experimental campaign, the following modifications are being introduced:

1) In order to avoid mirror deformations possibly caused by stress in the coating, a thinner titanium layer will be used on the optical surface (< 30 nm, or no coating). This should avoid unwanted crossing of the intensity lines in the intra-focal trace.
2) A more complex pattern of piezoelectric patches will be used, aiming at actuating a larger fraction of the mirror surface (Fig. 7), therefore increasing the actuation capabilities.
3) Piezoelectric patches of different sizes will be used in different geometries in 3 different mirror sector will be adopted to investigate the most effective solution.
4) The piezoelectric patches will be glued keeping the glass in contact with the integration mould, in order to minimize deformations.
5) The common electrode will be as thin as the others, in order to avoid mirror deformation induced by the stress in the metallic films used to deposit the contacts.
6) A system of two electronic boards will be used to drive 16 independent channels per mirror tested.
7) A removable connector will be used, instead of conductive glue, to connect the electronics to the contacts on the glass.
8) The active mirror will be mounted closer (1 m vs. almost 5 m used in the last tests) to the MCP at XACT to broaden the image and improve the resolution, reducing the possibility of reflected rays crossing each other.

We expect the next test at XACT to cast some light on the possibility to drive the shape of thin glass X-ray mirrors using the technique of intra-focal X-ray imaging.

## ACKNOWLEDGMENTS

The AXYOM project, devoted to the study of the correction of thin glass/plastic foils for X-ray mirrors, is financed by a TECNO-INAF 2012 grant. We thank Alexey Vikhlinin (Harvard-Smithsonian Center for Astrophysics) for useful discussions.